\documentclass{vak2024}
\usepackage[utf8]{inputenc}
\usepackage{url}
\usepackage{hyperref}

\begin{document}
\title{Regular distribution of star formation regions along the spiral arms and rings of disk galaxies}
\titlerunning{Regular distribution of SF regions}  
\author{A.~S.~Gusev\and E.~V.~Shimanovskaya}
\authorrunning{Gusev \& Shimanovskaya} 
	%
	%
\institute{Sternberg Astronomical Institute, Lomonosov Moscow State University, 13 Universitetsky Pr, Moscow 119234, Russia}

\abstract{
Last years studies have shown that the spatial regularity in the distribution of young stellar population along the spiral arms and rings of galaxies, previously considered to be rare, is a fairly common phenomenon. Spatial regularity has been found in the spiral arms and rings of galaxies of various morphology, from lenticular to extremely late-type spiral. The characteristic regularity scale is equal to 350-500~pc or a multiple thereof in all studied galaxies. Theoretical models predict a scale of instability of the stellar-gas disk on the order of a few kpc, which is several times larger than observed, although the most recent magneto-hydrodynamic simulations predict the formation of regular chains of star formation regions in spiral arms on a scale of 500-700 pc for the Milky Way-like galaxies. Modern high-quality surveys, such as PHANGS--MUSE, provide the necessary observational data (surface densities and velocity dispersions of gas and stellar population) to directly calculate the regularity scales in galaxies with high spatial resolution and wide field of view, which is a promising direction for research in this field.

\doi{10.26119/VAK2024-ZZZZ}
}

\maketitle

\section{Introduction}

\citet{elmegreen1983} first brought attention to the fact that neighboring HII regions in the spiral arms of some galaxies are located at equal distances from each other. They noted the rarity of this phenomenon: the only $\approx10\%$ of studied galaxies had visually detectable regular chains of HII regions with the characteristic distances $\lambda$ between adjacent HII regions of 1-4~kpc in different galaxies. Later, \citet{efremov2009} found a similar regularity in the  distribution of HI superclouds in the Carina and Cygnus spiral arms of the Milky Way with $\lambda=1.5$ and 1.3~kpc, accordingly. He also identified a regular chain of stellar complexes within the north-western arm of M31 with a characteristic distance of 1.1~kpc \citep{efremov2010}. 

The rarity of spatial regularities in the distribution of gas and stellar condensations in spiral arms of galaxies appears to be quite understandable. Theoretical studies of the gravitational instability of stellar-gas disks \citep{safronov1960,rafikov2001,romeo2013} reveal that the wavelength of instability depends on a comprehensive set of parameters, including gas and stellar surface densities and their velocity dispersions. The regularity in the distribution of gas clouds and their descendants, star formation regions, necessitates the constancy of these physical parameters in the stellar-gas disk within a wide range of galactocentric distances. However, such constancy should be a rare phenomenon in classical galactic disks.

Note that considering typical values of the parameters of the stellar and interstellar medium, all theoretical models predict the instability wavelength on the order of several kiloparsecs \citep{marchuk2018,inoue2021a}, which is in agreement with observational results of \citet{elmegreen1983} and \citet{efremov2009,efremov2010}.

However, the last decade studies of the regularity in the distribution of the young stellar population and gas clouds along the spiral arms and rings of galaxies have reshaped our understanding of the prevalence of this phenomenon and its characteristic spatial scales. The goal of this paper is to survey the latest observational and numerical modeling results of the patterns of spacial distribution of young stellar population and gas clouds along the spiral arms and rings in galaxies of various types. These results were obtained both by our own research team and by other scientific groups.

\section{Review of results}

\begin{figure*}
\centerline{\includegraphics[width=1.0\textwidth]{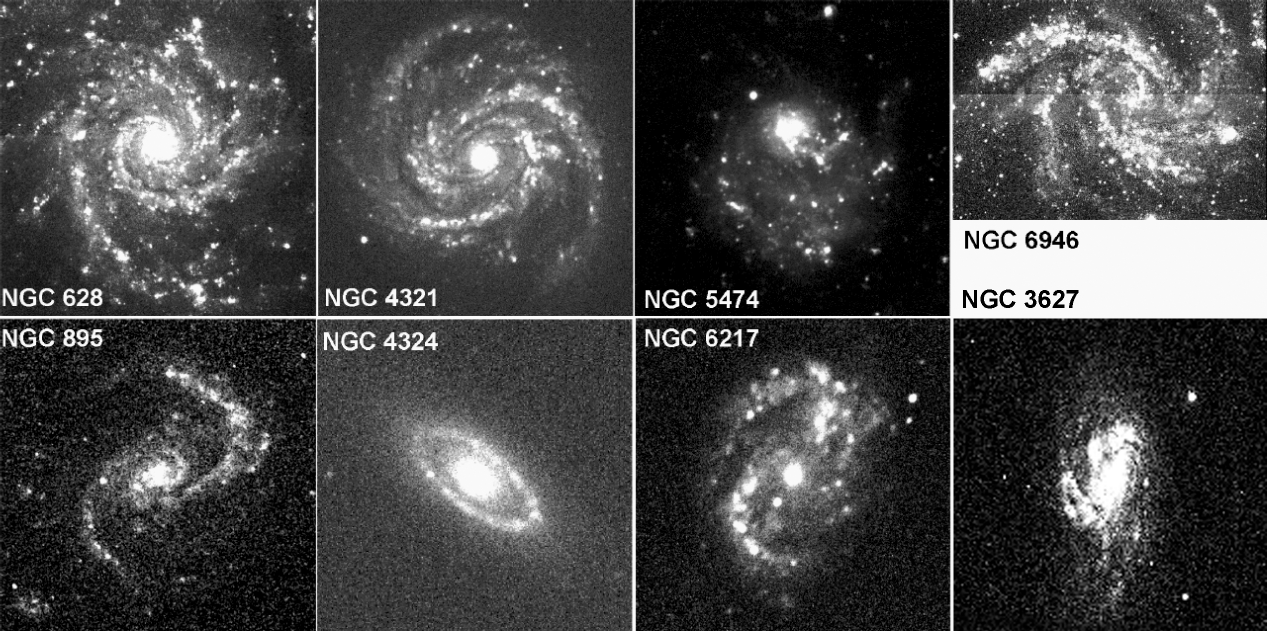}}
\caption{Images of galaxies with regular chains of star formation regions found in their spiral arms or rings. The images of NGC~3627 and NGC~4324 were obtained from the SDSS $u$ band; other galaxies are shown in the $U$ band. North is up, east is to the left.}
\label{fig:map}
\end{figure*}

Since 2013, the regular chains of HII regions, HI clouds, and star formation regions were identified in spiral arms of NGC~628 \citep{gusev2013}, NGC~4321 \citep{elmegreen2018}, NGC~895, NGC~5474, NGC~6946 \citep{gusev2022}, NGC~3627 (not previously published), in the Carina Arm of the Milky Way \citep{park2023}, in rings of NGC~6217 \citep{gusev2020} and NGC~4324 \citep{proshina2022}. The maps of these galaxies are presented in Fig.~\ref{fig:map}. The image of NGC~4321 in the figure obtained in \citet{marcum2001} was taken from the NED database\footnote{\url{http://ned.ipac.caltech.edu/}}; the images of NGC 3627 and NGC~4324 were given from the SDSS survey database\footnote{\url{http://www.sdss.org/dr13/}}; images of other galaxies are taken from the observations according to the author's programs \citep[see][]{gusev2013,gusev2020,gusev2022}.

\begin{figure}
\centerline{\includegraphics[width=1.0\textwidth]{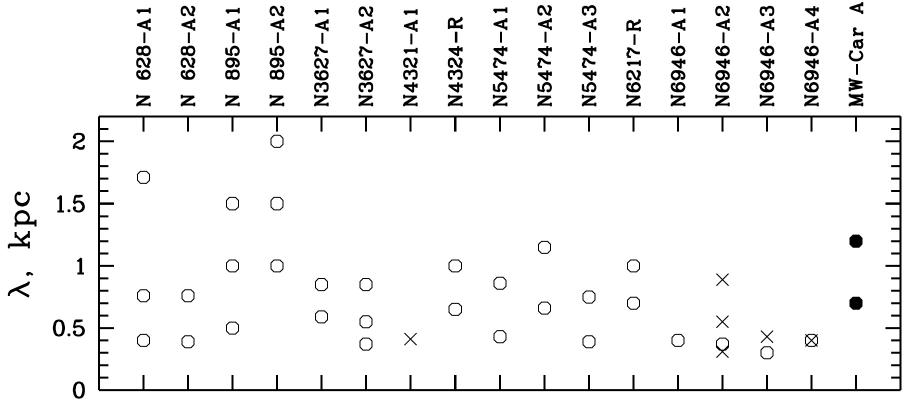}}
\caption{Characteristic distances $\lambda$ between neighboring local maxima of brightness in spiral arms and rings of different galaxies obtained using ultraviolet and optical (open circles), infrared (crosses), and radio data (filled circles).}
\label{fig:data}
\end{figure}

These results were obtained based on the analysis of images of galaxies in ultraviolet ({\it GALEX} FUV and NUV bands), optical ($UB$ bands and H$\alpha$ line), infrared ({\it Spitzer} 8$\mu$m) and radio (21~cm) ranges. We present the summary data on the characteristic distances between adjacent star formation regions (gas clouds, HII regions) in the spiral arms and rings of these galaxies in Fig.~\ref{fig:data}.

A detailed description of the technique and analysis of the results were given in the relevant papers and summarized in survey of \citet{gusev2023}. The only exception is NGC~3627, which has not been described previously. To study the regularities in this galaxy, we used its FUV image obtained from {\it GALEX} \citep{paz2007}, SDSS $u$ image \citep{brown2014}, and H$\alpha$ image from the {\it SIRTF Nearby Galaxies Survey (SINGS)}\footnote{\url{http://irsa.ipac.caltech.edu}} \citep{kennicutt2003}. All images were downloaded via NED database. 

\begin{figure}
\centerline{\includegraphics[width=1.0\textwidth]{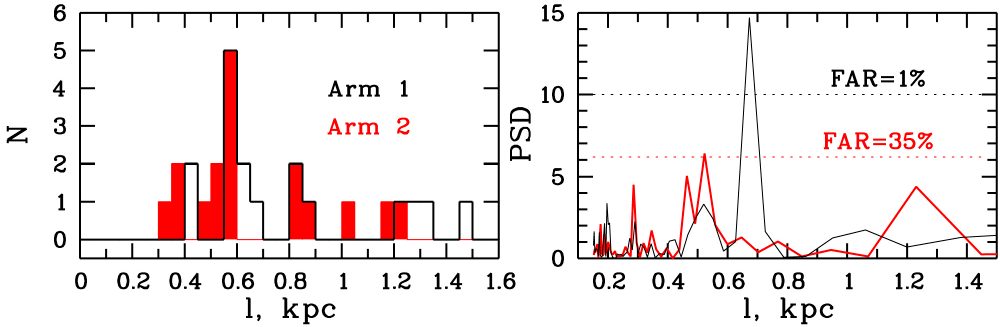}}
\caption{The number distribution histograms of local maxima of brightness by separation between adjacent young stellar condensations along the spiral arms of NGC~3627 (left) and normalized power spectral density of function $p(s)$ for the spiral arms in NGC~3627 (right). The function $p(s)$ is a collection of Gaussians, centred at points of local maxima of brightness on the profiles, with a dispersion equal to the peak positioning error; dotted lines are the false-alarm probability (FAP) levels of $1\%$ for Arm~1 and $35\%$ for Arm~2 \citep[see][for details]{gusev2022}.}
\label{fig:n3627}
\end{figure}

Using photometric profiles along the arms, we found 18 local brightness maxima (condensations of the young stellar population) in each of the spiral arms of NGC~3627. The results of the analysis of the distances between neighboring young stellar condensations are shown in Fig.~\ref{fig:n3627}. In both arms, the histogram shows a main peak at $590\pm30$~pc in the western arm (Arm~1) and $550\pm40$~pc in the eastern spiral arm (Arm~2). Half of the star formation regions in each spiral arm have a neighbor at a separation of $\sim550-600$~pc. Secondary peaks (at least three separations) are observed at $850\pm30$~pc in Arm~1 and $375\pm25$ and $850\pm25$~pc in Arm~2 (Fig.~\ref{fig:n3627}). The main peaks in the arms are also confirmed by the Fourier analysis data, which give a value of $670\pm20$~pc in Arm~1 and $525\pm20$~pc in Arm~2 (Fig.~\ref{fig:n3627}).

Summarizing the observational data, we can conclude that (i) regularity in the spatial distribution of young stellar population and gas clouds along the spiral arms and rings of galaxies of different morphology and structure is observed more often than expected (see Fig.~\ref{fig:map}); (ii) the characteristic separation between neighboring zones of concentration of the young stellar population is equal to or a multiple of 350-500~pc; (iii) brighter and larger stellar groupings (gas clouds) being located at greater (multiples) distances from each other than smaller and less massive \citep{gusev2013,park2023}; (iv) the presence or absence of shock waves does not affect the formation of regular chains of star formation regions along galactic spirals and rings \citep{gusev2013,gusev2020}.

Although theoretical models predicted a scale of instability of the stellar-gas disk on the order of a few kpc, recently \citet{arora2023} using magneto-hydrodynamic simulations in a galaxy like the Milky Way predicted the formation of regular chains of star formation regions in spiral arms on scales of $\sim500$~pc in the hydro case (without magnetic field) and $\sim650$~pc in the magnetic case.

\section{Discussion and conclusions}

Despite the simulation of \citet{arora2023}, the question of agreement between the observational data showing a scale of $\sim350-500$~pc and theoretical models (scale $>1$~kpc) remains open. We believe that a promising direction for research in this field is the use of modern high-quality surveys, such as PHANGS--MUSE, which provide the necessary observational data (parameters of gas and stellar population and kinematics of disk) for directly calculating regularity scales in galaxies with high spatial resolution and wide field of view.

At the same time, the question arises: how accurately do existing theoretical models predict the scale of fragmentation of gas clouds? Simulations of \citet{arora2023} and \citet{inoue2021b} showed that the separations obtained from numerical simulations turned out to be 1.5-2 times less than the wavelength of disk instability. A possible explanation for this discrepancy may be the difference between the local physical parameters of the gas disk at the moment of fragmentation from the observed ones. \citet{park2023} showed that if the observed HI clouds in Carina Arm formed by gravitational instabilities in a previously more uniform spiral arm gas, then the velocity dispersion $\sigma$ in this gas had to be less than or equal to $\sim4$~km~s$^{-1}$ whereas observed $\sigma=8-8.5$~km~s$^{-1}$. An alternative possibility is the higher gas surface density at the moment of fragmentation, caused by shock waves.

\acknowledgements{This research has made use of the NASA/IPAC Extragalactic Database (NED; {\url{http://ned.ipac.caltech.edu}}), SIRTF Nearby Galaxies Survey (SINGS; {\url{http://irsa.ipac.caltech.edu}}), the Sloan Digital Sky Survey (SDSS; {\url{http://www.sdss.org}}), and the Galaxy Evolution Explorer (GALEX; {\url{http://galex.stsci.edu}}) data.}


\end{document}